\documentclass[12pt]{article}
\usepackage{latexsym}
\newcommand{\bba}{\begin{eqnarray}}
\newcommand{\eea}{\end{eqnarray}}
\newcommand{\bb}{\begin{equation}}
\newcommand{\ee}{\end{equation}}
\newcommand{\bban}{\begin{eqnarray*}}
\newcommand{\eean}{\end{eqnarray*}}

\newcommand{\dddotV}{\buildrel{...}\over V}
\newcommand{\ddddotV}{\buildrel{....}\over V}
\begin{document}

\begin{titlepage}

\title{Stress-energy tensor in the
Bel-Szekeres space-time}

\author{\null}
\author{  {\sc Miquel  Dorca}\footnote{
E-mail: mdorca@rainbow.uchicago.edu}$\;\;$\footnote{
Present address: Box 1843, Department of Physics, 
               Brown University, Providence RI 02912
                                              } \\ 
          {\small\em  Enrico Fermi Institute}   \\
          {\small\em  The University of Chicago}  \\
          {\small\em  5640 Ellis Avenue}        \\
          {\small\em  Chicago IL 60637}    }

\date{February 17, 1998}

\maketitle

\begin{abstract}

In a recent work an approximation procedure was introduced to calculate the
vacuum expectation value of the stress-energy tensor
for a conformal massless scalar field in the classical background
determined by a particular colliding plane wave space-time.
This approximation procedure consists in appropriately modifying the
space-time geometry throughout the causal past of the collision center.
This modification in the geometry allows to simplify the boundary
conditions involved in the calculation of the Hadamard
function for the quantum state which represents the vacuum in the flat region
before the arrival of the waves. In the present work this approximation
procedure is applied to the non-singular Bel-Szekeres solution, which describes
the head on collision of two electromagnetic plane waves.
It is shown that the stress-energy tensor is unbounded as the 
killing-Cauchy horizon of the interaction is approached 
and its behavior coincides with
a previous calculation in  another
example of non-singular colliding plane wave space-time.

\end{abstract}



\vskip 0.5 truecm
\noindent
EFI-98-07

\end{titlepage}

\section{Introduction}

It is known that exact {\it gravitational plane waves}
are very simple time dependent plane symmetric solutions of Einstein's
equations \cite{bel26}. Nevertheless, they show two main nontrivial global 
features, namely:
i) the absence of a global Cauchy surface,
which is a consequence of the focusing effect that the waves exert on
null rays \cite{pen65},
ii) the presence of a Killing-Cauchy horizon which may be physically
understood as the caustic produced by the focusing of null rays \cite{pir89}. 
The inverse of
the focusing time is a measure of the strength of the wave. For an 
Einstein-Maxwell plane wave such inverse
time equals the electromagnetic energy per unit surface of the wave. This 
makes exact plane waves very
different from their linearized counterparts, which have no focusing points 
and admit a globally hyperbolic
space-time structure. One expects that exact plane waves may be relevant for 
the study of the strong time
dependent gravitational fields that may be produced in the collision of black 
holes \cite{dea79,fer80} or to
represent travelling waves on strongly gravitating cosmic strings 
\cite{gar89-90}. In recent years these
waves have been used in classical general relativity to test some conjectures 
on the stability of Cauchy
horizons
\cite{ori92,yur93}, and in string theory to test classical and quantum string 
behaviour in strong
gravitational fields \cite{veg84-90,veg91,jof94}. Their interest also stems 
from the fact that plane waves are a
subclass of exact classical solutions to string theory 
\cite{ama84-88,hor90,tse93fes94rus95}. 

In Einstein-Maxwell theory the particular class of plane symmetric
waves are seen to contain only a non-null component of the Ricci tensor and
only a non-null component of the Weyl tensor. In particular,
the single component of the Weyl tensor may be conveniently
interpreted as the {\em transverse wave component} in the direction of
propagation of the wave. In that sense,
the modulus term of the Weyl component can be identified with the
{\em amplitude} of the wave and the phase term with the {\em polarization}
of the wave. Furthermore, depending on whether the Ricci component or the Weyl
component is zero we will distinguish in between {\em pure gravitational
plane waves} or {\em pure electromagnetic plane waves} respectively.

When we consider a plane wave collision, we should analyze separately the
collision between pure gravitational waves,
between pure electromagnetic waves or between mixed waves. Namely:
i) when two pure gravitational
plane waves interact, the focusing effect of each 
wave distorts the causal structure of the space-time
near the null horizons that these waves contain and either a spacelike 
curvature singularity or a new regular Killing-Cauchy
horizon is created,
ii) when two pure electromagnetic plane waves interact, the situation
is more subtle. In fact,
in the full Einstein-Maxwell
theory, Maxwell's equations remain linear indicating non direct
electromagnetic interaction between the waves. However, 
there is a non-linear interaction of the waves through the gravitational 
field generated by their electromagnetic energy, which is
similar to the magnitude of the interaction between
pure gravitational waves. In that sense, the collision of two
electromagenic waves is seen to produce gravitational waves.
iii) in the case of mixed collisions, the pure electromagnetic wave is
partially reflected by the incident pure gravitational wave. The
gravitational wave, however, is not necessarily reflected. 

Note that, the presence of a Killing-Cauchy horizon in a colliding
plane wave space-time implies a breakdown of predictability since the
geometry beyond the horizon is not uniquely determined by the initial
data posed by the incoming colliding waves. Also, the singularities 
derived from plane wave collisions are not the 
result of the collapse of matter but the result of the non-linear
effects of pure gravity, i.e. the {\em self-gravitation} of the
gravitation field.

When waves are coupled to quantum fields there is
neither vacuum polarization nor the spontaneous creation of particles.
In that sense they behave 
very much as electromagnetic or
Yang-Mills plane waves in flat space-time \cite{des75,gib75}.
However, as a result of the non-linear interaction,
the creation of quantum particles is expected in a plane wave collision.

The interaction of quantum fields with colliding plane waves was first
considered by Yurtsever \cite{yur89} for the singular Khan-Penrose
solution \cite{kha71}, which describes the collision of two plane impulsive
gravitational waves. In that case,
an unambiguous ``out'' vacuum state was possible to define in a relatively
simple way. More recently, Dorca and Verdaguer \cite{dor93,dor94} 
noticed that the presence
of a Killing-Cauchy horizon in a non singular colliding plane wave 
space-time could be used to define an unambiguous ``out'' vacuum state
related to the preferred Hadamard state introduced by Kay and Wald in more
generic space-times with Killing-Cauchy horizons \cite{kay91}. 
With this premise,
Dorca and Verdaguer studied the interaction of quantum fields in a 
particular non-singular colliding plane wave space-time, the
interaction region of which was isometric to a region 
inside the event horizon of
a Schwarzschild black hole \cite{fer87cha86,hay89}. Later on, the same
premise was applied by Feinstein and Sebasti\'an \cite{fei95} 
to the Bel-Szekeres
solution \cite{bel74}, which represents the head 
on collision of two electromagnetic
plane waves with an interaction region isometric to the Bertotti-Robinson
universe \cite{rob54ber59} filled with an uniform electric field.
In all these examples it was found that the initial state, defined
to be the vacuum state in the flat region before the arrival of the waves,
contained a spectrum of ``out'' particles consistent, in the
long wavelength limit,  with a thermal spectrum with a temperature inversely
proportional to the focusing time of the waves.

A further step in the study of the interaction of quantum fields with
colliding plane waves is the computation of the expectation value of
the stress-energy tensor. Again, this problem was first considered by
Yurtsever \cite{yur89} for the Khan-Penrose solution \cite{kha71}.
In that case it was possible to determine the behavior of the stress-energy
tensor near the singularity of the interaction region.
It was shown that for the conformal coupling case (i.e. $\xi =1/6$) the
energy density and two of the principal pressures  were 
positive and unbounded towards the singularity. 
This problem has been also considered by Dorca and Verdaguer in 
the mentioned above non-singular colliding plane wave spacetime with an
interaction region isometric to an interior region of a Schwarzschild
black hole. As in the case of Yurtsever for
the Khan-Penrose solution, the expectation value of the stress-energy
tensor was calculated in the state representing the Minkowski vacuum
in the flat region before the arrival of the waves. This value
was first computed in a region close to both the Killing-Cauchy
horizon and the topological singularities, the {\em folding singularities}, 
that the colliding plane wave space-time contains \cite{gri91}.
In that particular region, the calculations were simplified due to
the  blueshift effect on the energy of the initial quantum modes as
they reached the Killing-Cauchy horizon \cite{dor96}. In a recent work
\cite{dor97}, an approximation procedure was proposed by the author
in order to calculate such an expectation value  throughout the causal
past region of the collision center. In both calculations, it was found that
the stress-energy diverged as the Killing-Cauchy horizon was approached.
The rest energy density was positive and unbounded towards the horizon.
Two of the principal pressures were negative and of the same order of 
magnitude of the energy density. It was also pointed out that such a
behavior suggested that the non singular Killing-Cauchy horizon is
indeed unstable under quantum perturbations and a curvature singularity
would be the general outcome of a generic plane wave space-time
when backreaction is taking into account.

In the present paper, the approximation procedure introduced in
\cite{dor97} is applied to the non singular Bel-Szekeres space-time as 
a first attempt to generalize such an approximation to more generic
colliding plane wave space-times.
The plan of the paper is the following. In section 2 the geometry of the 
Bel-Szekeres solution is
briefly reviewed. In section 3 an adequate approximation in the
space-time geometry is introduced throughout the causal past of the
collision center. Then, the mode solutions of a massless scalar field
which represent the vacuum state before the arrival of the waves
are calculated all over this particular region. In section 4 the
Hadamard function, which is the key ingredient for 
the computation of the stress-energy tensor, 
is calculated and regularized by means of
the {\em point-splitting} technique. In section 5 
the vacuum expectation value of
the stress-energy tensor is calculated. In section 6 a summary and 
some consequences of that result are given. In order to help to
maintain the main body of the paper reasonably clear, some final
results are stored in the Appendices.

\section{Description of the geometry}

The Bel-Szekeres solution \cite{bel74} represents the collision of two
electromagnetic shock waves followed by trailing radiation. The interaction
region is isometric to the Bertotti-Robinson universe
\cite{rob54ber59}, which is  the static conformaly flat solution of
Einstein-Maxwell equations with an uniform electric field.
Such a geometry is similar to the throat of the Reissner-Nordstrom
solution for the special case $M=Q$ \cite{whe73}.
The space-time
contains four space-time regions, given by

\bb \left.ds^2_{IV}\right.=4L_1L_2dudv-dx^2-dy^2,  \label{eq:ibIV}\ee
\bb \left.ds^2_{III}\right.=4L_1L_2dudv-
\cos ^2v\left( dx^2+ dy^2\right),\label{eq:ibIII}\ee
\bb \left.ds^2_{II}\right.=4L_1L_2dudv-
\cos ^2u\left(dx^2+dy^2\right),\label{eq:ibII}\ee
\bb \left.ds^2_{I}\right.=4L_1L_2dudv-
\cos^2 (u+v)\, dx^2-\cos ^2(u-v)\, dy^2,\label{eq:ibI}\ee
where for convenience we have used $u$ and $v$ as dimensionless
null coordinates, and where  $L_1$ and $L_2$, are length
parameters such that $L_1L_2$ is directly related to the focusing time
of the collision, i.e. to the inverse of the strength of the waves,
which is a measure of the amount of nonlinearity of the gravitational
waves \cite{gri91}.

This colliding wave space-time, as  shown in Fig. 1, consists of two
approaching waves, regions II and III, in a flat background, region
IV, and an interaction region, region I. The two waves move in the
direction of two null coordinates $u$ and $v$, and since they have
translational symmetry along the transversal $x$-$y$
planes, the interaction region retains a
two-parameter symmetry group of motions generated by the Killing
vectors $\partial _x$ and $\partial _y$.
The four space-time regions
are separated by the two null wave fronts $u=0$ and $v=0$. Namely, the
boundary between regions I and II is  $\{0\leq u<\pi /2,\; v=0\}$, the
boundary between regions I and III is $\{u=0,\; 0\leq v <\pi /2\}$, and
the boundary of regions II and III with region IV is
$\{u\leq 0,\;v=0\}\cup\{u=0,\;v\leq 0\}$. Region I meets region IV
only at the surface $u=v=0$. The Killing-Cauchy horizon in the region
I corresponds to the hypersurface $u+v=\pi /2$ and plane wave regions
II and III meet such a  Killing-Cauchy horizon
only at ${\cal P}=\{u=\pi /2,\; v=0\}$ and
${\cal P}'=\{u=0,\; v=\pi /2\}$ respectively. 
Observe that plane wave regions II
and III contain a singularity
at $u=\pi /2$, for region II, and $v=\pi /2$, for region III. These
singularities are not curvature singularities but a type of
topological singularity commonly referred to as a folding
singularity \cite{gri91}. This terminology arises from the fact that
the whole singularity $u=\pi /2$ in region II (or $v=\pi /2$ in region III)
must be identified (i.e. ``folded'') with $\cal P$ (or ${\cal P}'$)
(see \cite{dor93} for more details and for a 3-dimensional plot of a
space-time of this type).

\section{Mode propagation}

For simplicity we will consider in this section a massless scalar field,
which satisfies the usual Klein-Gordon equation,

\bb \Box\phi =0.\label{eq:kG}\ee
Following the directions of the approximation procedure introduced
in the previous work \cite{dor97}, 
we will be interested in the value of the quantum field
$\phi$ all over the causal past region of the collision center. 
The reason is essentially because the calculations can be greatly
simplified in this region. We will
start with the field solution in the flat region prior to the arrival
of the waves, which is chosen to be the usual vacuum in Minkowski
space-time. This vacuum solution will set a well posed initial value
problem on the null boundary $\Sigma =\{u=0,\; v\leq
0\}\cup\{u\leq 0,\; v=0\}$, by means of 
which a unique solution for the field equation can be found throughout
the space-time, i.e., in the plane wave regions (regions II and III), 
and in the interaction region (region I).

We will consider the line element,

\bb ds^2=2{\rm e}^{-M(u,v)}dudv-{\rm e}^{-U(u,v)}\left(
{\rm e}^{V(u,v)}dx^2+{\rm e}^{-V(u,v)}dy^2
\right), \label{eq:dsG}  \ee
which applies globally to the four space-time regions, and where the
functions $U$, $V$ and $M$, can be directly read off
(\ref{eq:ibIV})-(\ref{eq:ibI}). Then, the
field equation can
be separated in a plane-wave form solution for the transversal
coordinates $x$ and $y$, with $k_x$ and $k_y$, respectively, as
separation constants. This plane-wave separation is just a trivial 
consequence of the translational symmetry of the space-time on the planes
spanned by the Killing vectors $\partial _x$ and $\partial _y$. The
field solution is thus,

\bb \phi (u,v,x,y)={\rm e}^{U(u,v)/2}\, f(u,v)\, {\rm e}^{ik_xx+ik_yy},
\label{eq:phiG}\ee
where the function $f(u,v)$ satisfies the following second order
differential equation,

\bb f_{,uv}+\Omega (u,v)\, f=0;\;\;\; \Omega (u,v)=-
{\left({\rm e}^{-U/2}\right)_{,uv}\over {\rm e}^{-U/2}}+{1\over 2}
{{\rm e}^{-M+U}}\left( k_x^2{\rm e}^{-V}+k_y^2{\rm e}^V\right).
\label{eq:f(u,v)}\ee
Equation (\ref{eq:f(u,v)}) can be straightforwardly solved in the
flat region (region IV). Then, this solution determines 
on the null boundary $\Sigma$ a well posed set of
initial conditions for 
the solutions of equation (\ref{eq:f(u,v)}) in plane wave
regions II and III. Finally, the field solution in regions II,
III and IV is, 

\bb \phi (u,v,x,y)={1\over\sqrt{2k_-(2\pi)^3}}\,{\rm e}^{ik_xx+ik_yy}\,
\left\{\begin{array}{lll}
\displaystyle
{1\over\cos u}{\rm e}^{-i2{\hat k}_+\tan u-i2{\hat k}_-v};
& {\rm in\; region\; II}, & (a)\\
& &\\
\displaystyle
{1\over\cos v}{\rm e}^{-i2{\hat k}_+u-i2{\hat k}_-\tan v};
& {\rm in\; region\; III}, & (b)\\
& &\\
\displaystyle
{\rm e}^{-i2{\hat k}_+u-i2{\hat k}_-v};
& {\rm in\; region\; IV}, & (c)
\end{array}\right. \label{eq:phiII/III}\ee
where we have used  two new separation constants $k_\pm$, which
are related to the previous ones $k_x$ and $k_y$ by the relation
$4\, k_+k_-=k_x^2+k_y^2$. For convenience, we define also the
following set of dimensionless constants:

\bb {\hat k}_\pm = \sqrt{L_1L_2}\, k_\pm ,\;\;\;\; 
k_1 = \sqrt{L_1L_2}\, k_x,\;\;\;\; k_2 = \sqrt{L_1L_2}\, k_y.
\label{eq:kiL}\ee

Even though $\Sigma =\,\{(u=0,\,v<0\}\cup\{u<0,\,v=0\}$ is a 
null hypersurface, a well defined scalar product is given by (see 
\cite{dor93} for details),

\bb (\phi _1,\phi _2)=-i\int dx dy\left[
\int _{-\infty} ^0 \left.\left(\phi _1 
{\buildrel\leftrightarrow\over\partial}_u \phi
_2^*\right)\right|_{v=0}\, du +
\int _{-\infty} ^0 \left.\left(\phi _1 
{\buildrel\leftrightarrow\over\partial}_v \phi
_2^*\right)\right|_{u=0}\, dv\right].  \label{eq:scalarproduct}\ee
Notice that the initial modes
(\ref{eq:phiII/III}) are well normalized on the boundary $\Sigma$
between the flat region and the plane wave regions, and this means,
from general grounds, that they remain well normalized on the boundary
$\Sigma _{\rm I}=\,\{u=0,\, 0\leq v<\pi /2\}\cup\{0\leq u<\pi /2,\, v=0\}$
between the plane waves and the interaction region. Thus,
the Cauchy problem for the interaction region is now well posed.

However, although it has been rather easy to find the solution of the
field equation in regions II and III which smoothly matches with the
Minkowski vacuum, it turns out to be a difficult problem for the 
interaction region. Observe that the Cauchy data for the interaction
region is imposed by the field modes (\ref{eq:phiII/III}a,b) on the lines 
$\Sigma _{\rm I}=\,\{u=0,\, 0\leq v<\pi /2\}\cup\{0\leq u<\pi /2,\, v=0\}$,
which are characteristic lines for the differential 
equation (\ref{eq:f(u,v)}). Thus, the only independent Cauchy data are the
values of the function $f(u,v)$ on them. Furthermore, we are only 
interested in finding the field solution on the causal past of the
collision center, which is determined by the simple condition
$u=v$. Then, the only relevant Cauchy data lie on the segments,
${\bar\Sigma}_{\rm I}=\,\{u=0,\,0\leq v<\pi /4\}\cup\{0\leq u<\pi /4,\, v=0\}$
(see \cite{gara64} for details).

In order to solve this partial problem, we start with the following
change of coordinates,

\bb {t} =u+v,\;\;\; {z}=v-u, \label{eq:xieta}\ee
in  equation (\ref{eq:f(u,v)}) and we obtain,

\bb f_{,{t}{t}}-f_{,{z}{z}}+\Omega ({t} ,{z})f=0,  \label{eq:fxieta}\ee
where the term $\Omega ({t} ,{z})$, using (\ref{eq:kiL}), is given by

\bb \Omega ({t} ,{z})={k_1^2+(\sin ^2{t})/4\over\cos ^2{t}}+
{k_2^2-(\sin ^2{z})/4\over\cos ^2{z}}. \label{eq:Vxieta}\ee
From now on, we will denote the causal past region of the collision
center by ${\cal S}=\,\{0\leq {t}<\pi /2,\, -\pi /4<z<\pi /4\}$.
Observe that the behaviour of the variables ${t}$ and ${z}$ in equation
(\ref{eq:fxieta}) in region $\cal S$ is
very different. Since ${t}$ runs from $0$ to $\pi /2$ and ${z}$ runs
from $-\pi /4$ to $\pi /4$ in this region, the term 
(\ref{eq:Vxieta}) blows up as coordinate ${t}$ goes to $\pi /2$, but
is perfectly smooth over the entire range of coordinate
${z}$. This fact suggests that in the whole region $\cal S$
the physical results that we may expect are directly
related to the coordinate ${t}$ and we may not expect any physically
remarkable change if we take ${z} =0$ in equation (\ref{eq:fxieta}). 
However, if
we want to be consistent with such an approximation, we must also
modify the boundary conditions that lie on the line segments
${\bar\Sigma}_{\rm I}$. Since on the boundary ${\bar\Sigma}_{\rm I}$
we have that 
${t} =\pm{z}$ and coordinate ${t}$ runs from $-\pi /4$ to $\pi /4$,
we must also take ${t} ={z} =0$. This means that the boundary
conditions on ${\bar\Sigma}_{\rm I}$, given in (\ref{eq:phiII/III}a,b), reduce
in such an approximation to the flat boundary conditions given in 
(\ref{eq:phiII/III}c). Therefore we
change the mode propagation problem for the colliding wave space-time
into a rather simpler Schr\"odinger-type problem, which is
clear from Fig. 2, and which requires only that we find a solution
to equation (\ref{eq:fxieta}) with initial conditions given by the
Minkowski flat modes below the hypersurface
$\{{t} =0,\, -\pi /4 <{z}<\pi /4\}$. 
Recall, however,  that  
none of the discussion above is
applicable when a solution of equation
(\ref{eq:fxieta}) in the neighborhood of the folding singularities
${\cal P}$ and ${\cal P}'$ is required. This is not
only because in that case  both coordinates ${t} $ 
and ${z}$ take values near $\pi
/2$ and thus the potential term (\ref{eq:Vxieta}) is unbounded as
$z\rightarrow\pi /2$, but
also because the boundary conditions (\ref{eq:phiII/III}a,b) are also
unbounded as the folding singularities at ${t} =\pm{z} =\pi /2$ are approached.
In that case the mode propagation problem is much more complicated and
a more detailed discussion is required (see
\cite{dor93,dor94,dor96,fei95}) for details).

In fact, rather than relying on the discussed approximations in the
exact field equation (\ref{eq:fxieta}), it will be necessary to
rewrite a new field equation using an adequate approximation to the
space-time geometry throughout the causal past region of the collision
center. This is essentially because the process of renormalization involves
the subtraction of the infinite divergences that arise from the
formal definition of the stress-energy tensor, and these divergences
can be expressed as entirely geometric terms, which are independent
of any possible 
approximations in the field equation.

Approximating the field equation (\ref{eq:fxieta}) in the
causal past of the collision center by taking ${z} =0$ is essentially
equivalent to changing 
the line element, in the causal
past of the collision center, by a related line element
obtained from (\ref{eq:ibI}) by setting ${z} =0$, i.e.,

\bb
d{\hat s}^2_{\rm I}=L_1L_2\left(d{t} ^2-d{z} ^2\right)
-\cos ^2{t}\, dx^2-dy^2.
\label{eq:dshatIb}
\ee
We will suppose that the line element (\ref{eq:dshatIb})
applies all over the causal past of the collision center, not
only in the interaction region but also through the plane wave regions
II and III in the sense of Fig. 2.
The plane wave collision starts at $t=0$ but
to avoid smoothness problems derived from such an
approximation, we will suppose that (\ref{eq:dshatIb}) applies
exactly on a range $\epsilon <{t}<\pi /2$, for a certain
$\epsilon >0$. In the range
$0\leq {t}\leq\epsilon$, as described
below,  we will interpolate a line element 
which smoothly matches with the flat space at $t=0$.
Nevertheless, the details of this matching will not affect the main physical
features.

The
exact field equation for this approximate space-time is,

\bb
\left(\Box +\xi  R\right)\phi =0,
\label{eq:Appfieldeq}\ee
where it is necessary to consider a coupling curvature term in the field
equation because, although the exact space-time is a vacuum
solution, we have a bounded nonzero value for
$R$ in the approximated space-time.
In order to solve this new field equation, we start rewriting  the line
element (\ref{eq:dshatIb}) in the following general way,

\bb
ds^2=(f_1f_2f_3)\, d{{t} ^*}^2-\left(f_1f_2\over f_3\right)\, d{z} ^2
-\left(f_2f_3\over f_1\right)\, dx^2-
\left(f_1f_3\over f_2\right)\, dy^2,
\label{eq:dsgen}
\ee
where the $f_i$ are functions of coordinate ${t}$ alone, which
for values of $0<\epsilon <{t}<\pi /2$, can be
straightforwardly determined by direct comparison with
(\ref{eq:dshatIb}) as $f_1({t})=\sqrt{L_1L_2}$,
$f_2({t})=\sqrt{L_1L_2}\, \cos{t}$, $f_3({t})=\cos{t}$.
For values ${t}\leq 0$ we take $f_1({t})=f_2({t})=\sqrt{L_1L_2}$, 
$f_3({t})=1$, which correspond to their values in flat
space. Finally,  in
the interval $0\leq{t}\leq\epsilon$, we smoothly interpolate each  $f_i({t})$
($i=1,2,3$) between these values.
Also,
in order to prevent singularities in the field
equation, we conveniently reparametrize coordinate ${t}$, by ${t}
^*({t})$, as follows,

\bb
{d{t} ^*\over d{t}}={1\over f_3({t})}.
\label{eq:dxi*dxi}\ee
Now, we use the following ansatz for the field solutions,

\bb
\phi _k=h({t} ^*)\,{\rm e}^{ik_xx+ik_yy+ik_{z}{z}},
\label{eq:ansatz}\ee
where the plane wave factor in coordinates $x$, $y$ is related to the
translational symmetry of the space-time along  the transversal directions
$x$, $y$, and the plane wave factor in
coordinate ${z}$ is just a consequence of our approximation. 
Then equation (\ref{eq:Appfieldeq})
directly leads to the following Schr\"odinger-like differential
equation for the function $h({t} ^*)$,

\bb
h_{,{t} ^*{t} ^*}+\omega ^2({t} )\, h=0,\;\;\;\;
V({t})\equiv\omega ^2({t} )=f_0^2({t} )+f_1^2({t} )\, 
k_x^2+f_2^2({t} )\, k_y^2+
f_3^2({t} )\, k_z^2,
\label{eq:heq}\ee
where the function $f_0({t})$ stands for,

\bb
f_0^2({t} )=\left[f_1({t} )f_2({t} )f_3({t} )\right]\,\xi  R.
\label{eq:omegafi}\ee
Such differential equation can be WKB solved, essentially because the
short wavelength condition holds, i.e. $\omega
^{-1}d/d{t}^*\ln\omega\ll 1$. Observe that this condition reduces to
$(d{t}/d{t}^*)\, dV/d{t}\ll 2\,\omega ^3$, which becomes
particularly accurate when the Killing-Cauchy horizon is approached
since in that case $d{t}/d{t}^*=f_3({t})\rightarrow 0$. Therefore, the mode
solutions
$\phi _k$ which reduce to the flat mode solutions in the region prior
to the arrival of the waves, are

\bb
\phi _k={{\hat\omega}^{1/2}\over\sqrt{(2\pi)^32k_-W({t} )}}
{\rm e}^{ik_xx+ik_yy+ik_3{z}-i\int ^{{t} ^*}W(\zeta )d\zeta ^*},
\label{eq:phifi}\ee
where we denote ${\hat\omega}^2=k_1^2+k_2^2+k_3^2$
with $k_1=\sqrt{L_1L_2}\, k_x$, $k_2=\sqrt{L_1L_2}\, k_y$,
$k_3=k_z$ and where $W({t} )$ stands for an
adiabatic series in powers of the time-dependent frequency 
$\omega ({t} )$ of the 
modes and its derivatives. Up to adiabatic order four (i.e. up
to terms involving four derivatives of $\omega ({t})$) 
$W({t})$ it is given by,

\bb W({t} )=\omega  +{A_2\over\omega ^3} +{B_2\over\omega ^5} 
+{A_4\over\omega ^5}
 +{B_4\over\omega ^7} +{C_4\over\omega ^9} +{D_4\over\omega 
^{11}},\label{eq:Wxi}\ee
where, using the notation ${\dot V}\equiv dV/d{t}^*$,
 
\bb A_2=-{{\ddot V}\over 8},\;\;\;\; B_2={5\over 32}\,{{\dot 
V}^2},\label{eq:An}\ee
 
$$ A_4={{\ddddotV}\over 32},\;\;\;\; B_4=-
{28\,{\dot V}\,{\dddotV}+19\,{\ddot
V}^2\over 128},\;\;\;\; C_4={221\over 258}\,{{\dot V}^2\,{\ddot V}},\;\;\;\;
D_4=-{1105\over
2048}\,{{\dot V}^4},$$
and 
$A_n$, $B_n$, ... denote the $n$ adiabatic terms in
$W({t})$.
Up to adiabatic order zero it is simply
$W({t} )=\omega({t} )$. Observe the two following facts:

\noindent
(i) Near the horizon ${t}=\pi /2$ we have $W({t})\simeq\omega({t})$.
This is because the higher adiabatic corrections vanish at the horizon.

\noindent
(ii) In the flat region prior to the arrival of the waves we have
$W({t} )={\hat\omega} =\left(k_1^2+k_2^2+k_3^2\right)^{1/2}$. In that case,
since $f_3=1$, we can use (\ref{eq:dxi*dxi}) to set
$ {t} ^*={t}$, where without loss of generality we choose the
value ${t}^*=0$ at ${t=0}$.
Therefore, the mode solutions (\ref{eq:phifi}) in the flat region reduce to,

$$
\phi _k^{{\rm IV}}={1\over\sqrt{(2\pi)^32k_-}}{\rm
e}^{ik_xx+ik_yy+ik_{z}{z} -i{\hat\omega}{t}},
$$
which indeed are the flat mode solutions defined in (\ref{eq:phiII/III}c),
recalling that the new separation constant $k_{z}=k_3$ is related to the
original $k_\pm$ by the ordinary null momentum relations, i.e.,

\bb
{\hat\omega}={\hat k}_++{\hat k}_-,\;\;{k_{z}}={\hat k}_+-{\hat k}_-.
\label{eq:nullmom}\ee

It is important to understand that we are constructing a set of mode
solutions as an adiabatic series in terms of derivatives of the
frequency $\omega ({t})$ in the differential equation
(\ref{eq:heq}). This procedure is similar but not equivalent to the
construction of an {\em adiabatic vacuum state} where the field modes
are expanded as an adiabatic series in terms of the derivatives of the
metric coefficients (see for example \cite{bir82} for details). In
fact, observe for instance that the term $f^2_0({t})$ in (\ref{eq:heq})
involves two derivatives of the metric since it is directly related to
the curvature scalar. Thus, it would be a second order term for 
an eventual adiabatic vacuum, but it is simply a zeroth order term in 
our adiabatic series in derivatives of $\omega ({t})$.

\section{Hadamard function in the interaction region}

The key ingredient to calculate the vacuum expectation value of the
stress-energy tensor is the {\em Hadamard function} $G^{(1)}(x,x')$,
which is defined as the vacuum expectation value of the anticommutator
of the field, i.e.,

\bb G^{(1)}(x,x')=\langle\{\phi (x),\phi (x')\}\rangle =\sum _k 
\left\{u_{k}(x)\, u^*_{k}(x')+u_{k}(x')\,
u^*_{k}(x)\right\}. \label{eq:defHadamard}\ee
This Hadamard function contains non-physical divergence terms which
can be subtracted by the following point-splitting prescription,

\bb G_B^{(1)}(x,x')=G^{(1)}(x,x')-S(x,x'),
\label{eq:GB}\ee
where $S(x,x')$ is the {\em midpoint expansion} of a locally 
constructed quantity commonly referred as
a {\em Hadamard elementary solution} 
(see for example \cite{wal94}) and given by

\bba
S(x,x')&=&{1\over 8\pi ^2}\left\{
-{2\over\sigma}
-2{\Delta ^{(2)}}_{{\bar\mu}{\bar\nu}}\,{\sigma ^{\bar\mu}\sigma
^{\bar\nu}\over\sigma}
-2{\Delta ^{(4)}}_{{\bar\mu}{\bar\nu}{\bar\rho}{\bar\tau}}
\,{\sigma ^{\bar\mu}\sigma ^{\bar\nu}\sigma ^{\bar\rho}\sigma
^{\bar\tau}\over\sigma}
 -a_1^{(0)}\,\ln (\mu ^{-2}\sigma ) 
\right.\nonumber
\\
& & \left.
-\left[
\left(
a_1^{(0)}{\Delta ^{(2)}}_{{\bar\mu}{\bar\nu}}
+{a_1 ^{(2)}}_{{\bar\mu}{\bar\nu}}
\right)\sigma ^{\bar\mu}\sigma ^{\bar\nu}
-{1\over 2}a_2^{(0)}\sigma
\right]\,\ln (\mu ^{-2}\sigma )
-{3\over 4}a_2^{(0)}\sigma
\right\},
\label{eq:Hada1}\eea
where the coefficients ${\Delta ^{(2)}}_{{\bar\mu}{\bar\nu}}$,
${\Delta ^{(4)}}_{{\bar\mu}{\bar\nu}{\bar\rho}{\bar\tau}}$,
$a_1^{(0)}\cdots$ are written
in Appendix A. We use the standard definition for the {\em geodetic
biscalar} $\sigma (x,x')=(1/2)s^2(x,x')$, being $s(x,x')$ the proper
distance between the points $x$ and $x'$ on a non-null geodesic
connecting them. Also, 
$\sigma _{\bar\mu} (x,x') =(\partial /\partial x^{\bar\mu})\sigma (x,x')$ 
is a geodesic tangent vector at the point $\bar x$ with modulus $s(x,x')$,
being $\bar x$ the {\em midpoint} between $x$ and $x'$ on the geodesic.
The parameter $\mu$ in the
logarithmic term of (\ref{eq:Hada1}) is an arbitrary length parameter,
which is related to the two-parameter ambiguity of the point-splitting 
regularization scheme \cite{wal94}.
Then we can compute $\langle T_{\mu\nu} \rangle$ by means of the
following differential operation,

\bb \langle T_{\mu\nu}(x)\rangle =\lim _{x\rightarrow x'}\, {\cal 
D}_{\mu\nu}G^{(1)}(x,x'), \label{eq:limDT}\ee
where ${\cal D}_{\mu\nu}$ is a nonlocal differential operator, which
in the conformal coupling case ($\xi =1/6$) is given by,

\bba{\cal D}_{\mu\nu}&=&
{1\over 6}\,\left(\nabla _{\mu '}\nabla _{\nu}+\nabla
_{\nu'}\nabla_{\mu}\right)
-{1\over 24}\, g_{\mu\nu}\,\left(
\nabla _{\alpha '}\nabla ^{\alpha}+
\nabla _{\alpha }\nabla ^{\alpha '}\right)-\nonumber\\
& &{1\over 12}\,\left(\nabla _{\mu }\nabla _{\nu}+\nabla
_{\mu'}\nabla_{\nu '}\right)
+ {1\over 48}\, g_{\mu\nu}\,\left(
\nabla _{\alpha  }\nabla ^{\alpha  }+
\nabla _{\alpha '}\nabla ^{\alpha '}\right)-\nonumber\\
& &{1\over 12}\,\left(R_{\mu\nu}-{1\over 4}R\, g_{\mu\nu}\right).
\label{eq:Dopdif}\eea
However, the above differential operation and its limit have no immediate 
covariant meaning because
$G^{(1)}(x,x')$ is not an ordinary function but a {\it biscalar} and the 
differential operator ${\cal
D}_{\mu\nu}$ is {\it nonlocal}; thus we need to deal with the nonlocal 
formalism of {\it bitensors} (see, for example \cite{dew60,chr76} or the
Appendix B of reference \cite{dor96} for a review on
this subject).

The regularization procedure (\ref{eq:GB}), however, 
fails to give a covariantly conserved 
stress-energy tensor essentially because the locally constructed
Hadamard function (\ref{eq:Hada1}) is not in
general symmetric on the endpoints $x$ and $x'$ (i.e. it satisfies the
field equation at the point $x$ but fails to satisfy it at $x'$)
(see \cite{wal78} for details).
Thus, to ensure covariant conservation, we must introduce an additional 
prescription:
 
\bb \langle T_{\mu\nu}(x)\rangle =\langle T_{\mu\nu}^B (x)\rangle
- {a^{(0)}_2(x)\over 64\pi ^2}\, g_{\mu\nu}. \label{eq:GBT}\ee
Note that this last term is responsible for the trace anomaly in the conformal 
coupling case, because even though
$\langle T_{\mu\nu}^B (x)\rangle$ has null trace when $\xi =1/6$, the trace of 
$\langle T_{\mu\nu}(x)\rangle$
is given by
$\langle T^{\mu}_{\mu}\rangle =
- {a^{(0)}_2(x)/(16\pi ^2)}$.
The regularization prescription just given in (\ref{eq:GBT})
satisfies the well known four  
Wald's axioms
\cite{wal94,wal76,wal77-78b,chr75}, a set of properties that any physically 
reasonable expectation value of the
stress-energy tensor of a quantum field should satisfy. There is still an 
ambiguity in this prescription since
two independent conserved local curvature terms, which are quadratic in the 
curvature, can be added to this
stress-energy tensor. In particular, the $\mu$-parameter ambiguity in 
(\ref{eq:Hada1}) is a consequence of this (see \cite{wal94} for details).
Such a two-parameter ambiguity, however, cannot be 
resolved within the limits of the
semiclassical theory, it may be resolved in a complete quantum theory
of gravity 
\cite{wal94}. Note, however, that in some sense this ambiguity does not affect 
the knowledge of the matter
distribution because a tensor of this kind belongs properly to the left hand 
side of Einstein equations, i.e.
to the geometry rather than to the matter distribution. 

After this preliminary introduction on the point-splitting regularization
technique, we may proceed to  calculate the Hadamard function $G^{(1)}(x,x')$ 
in the interaction region for the initial vacuum state
defined by the modes $\phi _k$, (\ref{eq:phifi}). The 
Hadamard function can be written as,
 
\bb G^{(1)}(x,x')=\sum _k\phi
_k(x)\,\phi ^*_k(x')\; +{c.c.}    
\label{eq:G(1)}\ee
Note that solutions $\phi _k$ contain the function $h({t} ^*)$, which
cannot be calculated analytically but may be approximated
up to any adiabatic order as described in (\ref{eq:Wxi})-(\ref{eq:An}). 
Thus, we have the 
inherent ambiguity of where to
cut the adiabatic series. In fact, this is an asymptotic expansion,
which has a well stablished ultraviolet limit but it may have convergence
problems in the low-energy limit.
However, observe from (\ref{eq:dxi*dxi}) and (\ref{eq:heq})
that since $dt/dt^*\rightarrow 0$ and
$V({t})=\omega ^2({t})\rightarrow k_1^2$ 
towards the horizon, the adiabatic series (\ref{eq:Wxi})
reduces to $W\simeq\omega$ near the horizon. This means that we could cut the 
adiabatic series (\ref{eq:Wxi}) at order zero if we were
interested in a calculation near the horizon. However, this is only
partially true. In fact, it would be true if we were only interested in the 
particle production problem \cite{fei95} but it is not
sufficient for the calculation of the vacuum expectation value of the
stress-energy tensor.
This is because
$G^{(1)}$ calculated with $h({t}^*)$
at order zero does not reproduce the short-distance singular 
structure of a Hadamard elementary
solution (\ref{eq:Hada1}) in the coincidence limit $x\rightarrow x'$. The 
smallest adiabatic
order for the function $h({t}^*)$ 
which we need to recover the singular structure of $G^{(1)}$ is order 
four, basically  because our adiabatic construction of the mode
solutions is similar (but not equivalent) to an {\em adiabatic vacuum
state} (see \cite{bir82} for details).
 
Although expanding the function $h({t}^*)$ in (\ref{eq:G(1)}) up to 
adiabatic order four will give an accurate value for
the stress-energy tensor near the horizon, it will also give a suitable
approximate value for this tensor 
all over the causal past of the collision center
(region $\cal S$ in Fig. 2). The reason is that even though the 
short-wavelength condition, i.e.
$\omega ^{-1}d/d{t}^*\ln\omega\ll 1$, is particularly accurate near the
horizon it also holds throughout region $\cal S$.

In the mode sum (\ref{eq:G(1)}) we use the shortened notation 
$\sum 
_k\equiv 
\int ^{\infty}_{0}{dk_-/ k_-}\,
\int ^{\infty}_{-\infty}dk_x\,
\int ^{\infty}_{-\infty}dk_y$ or equivalently
$\sum _k\equiv
(L_1L_2)^{-1}
\int ^{\infty}_{-\infty}dk_1\,
\int ^{\infty}_{-\infty}dk_2\,
\int ^{\infty}_{-\infty}{dk_3/ {\hat\omega}}$,
where the change of variables
(\ref{eq:nullmom}) and the usual notation (\ref{eq:kiL}) have been used.
Therefore we have, 
            
\bba  G^{(1)}(x,x')&=& {1\over 2(2\pi)^3\, L_1L_2}\,
\int ^{\infty}_{-\infty}\int ^{\infty}_{-\infty}\int
^{\infty}_{-\infty}
{dk_1\, dk_2\, dk_3\over \sqrt{W({t})W({t} ')}}\,\times \nonumber\\
& &{\rm e}^{-i\int _{{t^*} '}^{{t^*}}W({\zeta})d{\zeta}^*+ik_x
(x-x')+ik_y (y-y')+ 
ik_{z} ({z} -{z} ')}\;\; +c.c.                         
\label{eq:G(1)2}\eea
We assume that the points 
$x$ and $x'$ are connected by a
non-null geodesic in such a way that they are at the same proper distance 
$\epsilon$ from a third midpoint
$\bar x$. We parametrize the geodesic by its proper distance $\tau$ and 
with abuse of notation we denote 
the end points by
$x$ and $x'$, which should not be confused with the third
component of $({t} ,\;{z} ,\; x,\; y)$. 
Then we expand the integrand
function in powers of $\epsilon$ and we finally integrate term by term
to get an expression up to $\epsilon ^2$. The details of such a tedious
calculation can be found in ref. \cite{dor97}. The result is,

\bba G^{(1)}(x,x')&=&{\bar A}+\sigma\,{\bar b}
+{C}_{{\bar \alpha}{\bar \beta}}\,{\sigma}^{\bar \alpha}{\sigma}^{\bar\beta}
+{D}_{{\bar \alpha}{\bar \beta}{\bar \gamma}{\bar \delta}}\,
{\sigma}^{\bar \alpha}{\sigma}^{\bar\beta}{\sigma}^{\bar
\gamma}{\sigma}^{\bar\delta}
+{1\over 8\pi ^2}\left\{
-{2\over\sigma}
-2{\Delta ^{(2)}}_{{\bar\mu}{\bar\nu}}\,{\sigma ^{\bar\mu}\sigma
^{\bar\nu}\over\sigma}
\right.
\label{eq:Hada2}\\
& & \left.
-2{\Delta ^{(4)}}_{{\bar\mu}{\bar\nu}{\bar\rho}{\bar\tau}}
\,{\sigma ^{\bar\mu}\sigma ^{\bar\nu}\sigma ^{\bar\rho}\sigma
^{\bar\tau}\over\sigma}
-a_1^{(0)}\, L
-\left[
\left(
a_1^{(0)}{\Delta ^{(2)}}_{{\bar\mu}{\bar\nu}}
+{a_1 ^{(2)}}_{{\bar\mu}{\bar\nu}}
\right)\sigma ^{\bar\mu}\sigma ^{\bar\nu}
-{1\over 2}a_2^{(0)}\sigma
\right]\, L
\right\}
\nonumber\eea
where $L$ is a logarithmic term defined as 
$L=2\gamma+\ln (\sigma\,\xi R/2)$,
being $\gamma$ Euler's constant and  where all the involved
coefficients
$\bar A$, $\bar b$, ${C}_{{\bar \alpha}{\bar \beta}}$... are given in 
Appendix A.
 
According to (\ref{eq:GB}), the Hadamard function can be regularized using the 
elementary
Hadamard solution (\ref{eq:Hada1}) and finally the regularized expression for
$G^{(1)}(x,x')$ up to order $\epsilon ^2$ is,
 
\bba G^{(1)}_B(x,x')&=&{\bar A}+\sigma\,{\bar B}
+{C}_{{\bar \alpha}{\bar \beta}}\,{\sigma}^{\bar \alpha}{\sigma}^{\bar \beta}
+{D}_{{\bar \alpha}{\bar \beta}{\bar \gamma}{\bar \delta}}\,
{\sigma}^{\bar \alpha}{\sigma}^{\bar\beta}{\sigma}^{\bar
\gamma}{\sigma}^{\bar\delta}
\nonumber\\
& &+{1\over 8\pi ^2}\left\{
-a_1^{(0)}\, {\hat L}
-\left[
\left(
a_1^{(0)}{\Delta ^{(2)}}_{{\bar\mu}{\bar\nu}}
+{a_1 ^{(2)}}_{{\bar\mu}{\bar\nu}}
\right)\sigma ^{\bar\mu}\sigma ^{\bar\nu}
-{1\over 2}a_2^{(0)}\sigma
\right]\, {\hat L}
\right\},
\label{eq:G(1)fR}\eea
where ${\hat L}$ is a bounded logarithmic term given by,
${\hat L}=2\gamma+\ln (\mu ^{2}\,\xi R/2)$, $\mu$ being the arbitrary 
length parameter introduced in (\ref{eq:Hada1}),
and where the coefficient $\bar B$ is 
${\bar B}=b+3\, a_2^{(0)}/(32\pi ^2)$, which is given also in Appendix A.
From (\ref{eq:G(1)fR}) we can directly read off the regularized mean
square field in  
the ``in" vacuum state as
$\langle\phi ^2\rangle ={\bar A}/2-a_1^{(0)}{\hat L}/(16\pi ^2)$.
It is important to remark, however, that the term 
${D}_{{\bar \alpha}{\bar \beta}{\bar \gamma}{\bar \delta}}$ in
(\ref{eq:G(1)fR}) 
appears only as a consequence of our approximate procedure of
calculating the Hadamard function, i.e.  using an adiabatic order four
expansion for the initial modes in powers of the 
mode frequency $\omega ({t})$ and its derivatives.
Had we used an exact expression for the
initial modes (or an adiabatic vacuum state \cite{bir82}), such
a term would not appear.

\section{Expectation value of the stress-energy tensor}

To calculate the vacuum expectation value of the stress-energy
tensor we have to apply the differential operator 
(\ref{eq:Dopdif}) to
(\ref{eq:G(1)fR}). As we have already pointed out, this is not straightforward 
because we
work with nonlocal quantities. Note first that the operator (\ref{eq:Dopdif}) 
acts on bitensors which depend on the end points
$x$ and $x'$, but the expression (\ref{eq:G(1)fR}) for $G^{(1)}_B$ depends on 
the midpoint $\bar x$. This means
that we need to covariantly expand (\ref{eq:G(1)fR}) in terms of 
the endpoints $x$ and $x'$. Also, the presence of quartic $\sigma ^\mu$
terms in (\ref{eq:G(1)fR}) gives, after differentiation, path dependent
terms which must be conveniently averaged.
The details of such a calculation
may be found for instance in \cite{dor96,dor97}.
Then, in
the orthonormal basis 
$\theta _1={g^{1/2}_{{t}{t}}}\,d{t}$,
$\theta _2={g^{1/2}_{{z}{z}}}\, d{z}$, $\theta _3={g^{1/2}_{xx}}\, dx$, 
$\theta _4={g^{1/2}_{yy}}\, dy$,  using the trace anomaly
prescription (\ref{eq:GBT}),  we obtain
the following  expectation values $\langle T_{\mu\nu}\rangle$ in
the conformal coupling case and for values
$0<\epsilon<{t}<\pi /2$ of coordinate ${t}$
 
\bba\langle T_{\mu\nu}\rangle &=&
{2\gamma+\ln (\mu ^{2}\, R/12)\over 2880\, (L_1L_2)^2\, 
\pi ^2}\,{\rm diag}(-1,\, -1,\, 1,\, -1)
+{1\over 4}\langle T^{\tau}_{\tau}\rangle g_{\mu\nu}+
\nonumber
\\
&&
{\rm 
diag}\left(\rho _1({t}) ,\; -\rho _2({t}) ,\; \rho _1({t})+2\rho
_2({t}),\; -\rho _2({t})\right),
\label{eq:TMNconforme}\eea
where we have used for simplicity the
notation $z=\sin{t}$. 
The trace anomaly term in that case is given
by  (\ref{eq:GBT}) as,

$${1\over 4}\langle T^{\tau}_{\tau}\rangle g_{\mu\nu} =
-{a_2\over 64\,\pi ^2}\, g_{\mu\nu}=
-{{\rm diag}\left(1 ,\, -1 ,\, -1,\, -1\right)
\over 2880\, (L_1L_2)^2\, \pi ^2},
$$ 
and the functions $\rho _1({t})$ and $\rho _2({t})$
are  given by,
 
\bb\rho _1({t}) ={-247247+1456234\, z^2+792789\, z^4\over
69189120\, (L_1L_2)^2\,\pi ^2\,(1-z)^2(1+z)^2},\label{eq:rho1}\ee
\bb\rho _2({t}) ={502931+179686\, z^2+923887\, z^4\over 
69189120\, (L_1L_2)^2\, \pi ^2\,(1-z)^2(1+z)^2}.\label{eq:rho2}\ee
Recall that $\rho _1({t})$ is a positive definite function in an
interval $0<\epsilon <{t}<\pi /2$. Both functions are unbounded at the horizon
$({t}=\pi /2)$, and the expectation value of the stress-energy tensor
near the horizon is approximately given by,

\bb\left.\langle T_{\mu\nu}\rangle\right|_{{t}\simeq\pi /2} ={\rm 
diag}\left(\rho ({t}) ,\; -\rho ({t}) ,\; 3\rho ({t})
,\; -\rho ({t})\right)+
{1\over 4}\langle T^{\tau}_{\tau}\rangle g_{\mu\nu},
\label{eq:TMapprox}\ee
where $\rho ({t})=\Lambda (L_1L_2)^{-2}\cos{t}^{-4}$ and $\Lambda\simeq
0.0029$. The behavior of $\langle T_{\mu\nu}\rangle$ entirely agrees
with the previous result \cite{dor97}.
For values of $t\leq 0$, $\langle T_{\mu\nu}\rangle =0$, 
and to be consistent with the approximation 
we have used for the space-time geometry, we should
require that the value of $\langle T_{\mu\nu}\rangle$
(\ref{eq:TMNconforme}), which is valid for $0<\epsilon <{t}<\pi /2$, goes 
smoothly to zero as ${t}\rightarrow 0$. In fact, this can be achieved
using an adequate matching of the line element (\ref{eq:dshatIb}) with the
flat line element through the interval $0\leq{t}\leq\epsilon$. 

Observe that the logarithmic term in the stress-energy 
tensor (\ref{eq:TMNconforme}) appears as
a consequence of a similar term in the Hadamard function
(\ref{eq:G(1)fR}). The argument of this 
logarithm depends on the curvature scalar
and thus it will grow unbounded as the flat region is approached.
However, the coefficient that will appear in front  of such a logarithm
in the Hadamard function (\ref{eq:G(1)fR}),
depends only on locally constructed curvature terms. Therefore,
with an adequate matching of the space-time
geometry, this coefficient will also
smoothly vanish towards the flat space region, below $t=0$, and it
will not give a contribution to the stress-energy tensor.
The details of such a matching, however, will not affect
the main features of the stress-energy tensor (\ref{eq:TMNconforme}),
particularly when the Killing-Cauchy horizon is approached.

We must recall, however, that
although the value (\ref{eq:TMNconforme}) for 
$\langle T_{\mu\nu}\rangle$ satisfies asymptotically 
the conservation equation near
the Killing-Cauchy horizon, it does not satisfy exactly the conservation
equation throughout region $\cal S$, essentially
because it is obtained by means of an
approximation in the field modes. Nevertheless, we could obtain a
truly conserved $\langle T_{\mu\nu}\rangle$, in the context of the
present approximation, by solving the conservation equation
considering a $\langle {T^{\mu}}_{\nu}\rangle$ given by
the following set of non-null components
$\{\langle T^t_t({t})\rangle ,\, \langle T^z_z({t})\rangle ,\,
\langle T^x_x({t})\rangle ,\, \langle T^y_y({t})\rangle\}$ with the
conditions: 
i) $\langle T^y_y({t})\rangle =\langle T^z_z({t})\rangle$, 
which is compatible with
(\ref{eq:TMNconforme}) and it is a physical consequence of the isotropy
of the metric (\ref{eq:dshatIb}) along the $y$-$z$ directions,
ii) trace anomaly condition, i.e. 
$\langle T^x_x({t})\rangle =\langle T^{\mu}_{\mu}\rangle
-\langle T^t_t({t})\rangle -2\, \langle T^z_z({t})\rangle$, 
iii) the ansatz $\langle T^t_t({t})\rangle =\rho _1({t})$, 
which is the approximate value of $\langle T^t_t({t})\rangle $ obtained in
our calculation. Finally, the conservation equation gives
straightforwardly  values for 
$\langle T^z_z({t})\rangle$ and $\langle T^x_x({t})\rangle$
which are compatible with the values 
$\langle T^z_z({t})\rangle =\rho _2({t})$ and
$\langle T^x_x({t})\rangle =-\rho _1({t})-2\rho _2({t})$ obtained in our
approximation. In particular, they have the same behaviour near the
Killing-Cauchy horizon.
 
Inspection of (\ref{eq:TMNconforme}) shows that not only is 
the {\em weak energy
condition} satisfied \cite{wal84}, which means that the energy density is 
nonnegative for any observer, but also
the {\em strong energy condition} is satisfied.

\section{Conclusions}
 
We have calculated the expectation value of the stress-energy tensor
of a massless scalar field in a space-time representing the head on
collision of two electromagnetic plane waves throughout the causal
past of the collision center and in the field state which
corresponds to the physical vacuum state before the collision takes
place. We have performed the calculations in this particular region
essentially because, following the directions of a previous work
\cite{dor97}, we could introduce a suitable approximation to the
space-time metric (see Fig. 2). This approximation not only has allowed us to
dramatically simplify the calculations but also to keep unchanged the
main physical features, in particular the behavior of the
stress-energy tensor near the Killing-Cauchy horizon of the
interaction region. In fact, such an approximation is also valid for more
generic plane wave space-times, and this will be the subject of a
forthcoming paper.

The results we have obtained are 
entirely compatible with the previous result \cite{dor97}
and they may be briefly described as follows: before the collision
of the waves $\langle T_{\mu\nu}\rangle =0$, which correspond to the
lower edge Fig. 2. Then, after the collision the value of 
$\langle T_{\mu\nu}\rangle$ starts to increase until it grows
unbounded towards the Killing-Cauchy horizon of the interaction
region. The 
weak energy condition is satisfied, the rest energy density is positive and 
diverges as $\cos ^{-4}{t}$. Two of 
the principal pressures are negative and of the same order of magnitude 
of the energy density. The
strong energy condition is also satisfied, 
$\langle T^{\mu}_{\mu}\rangle$ is finite but  $\langle
T^{\mu\nu}\rangle \langle T_{\mu\nu}\rangle$ diverges at the horizon and we 
may
use ref \cite{hel93} on the stability of Cauchy horizons to argue that the 
horizon will acquire by backreaction
a curvature singularity.
Thus, contrary to  simple plane waves,
which do not polarize the vacuum \cite{des75,gib75}, the nonlinear collision 
of these waves polarize the vacuum
and the focusing effect that the waves exert seems to produce,
in general, an unbounded positive energy density at the focusing points.
Therefore, when the colliding waves produce a Killing-Cauchy horizon, that 
horizon may be, in general, unstable by vacuum
polarization.
 
In the more generic case when the wave collision produces a spacelike 
singularity it seems clear that the
vacuum expectation value of the stress-energy tensor will also grow unbounded 
near the singularity. In fact, in a forthcoming paper we will extend
the approximation introduced in the present work to a more generic
plane wave spacetime with the objective of more generally proving that the
negative pressures associated to the quantum fields could not prevent
the formation of the singularity.

\vskip 1.25 truecm
 
{\Large{\bf Acknowledgements}}
 
\vskip 0.5 truecm
 
\noindent
I am  grateful to R. M. Wald, R. Geroch, E. Verdaguer, A. Campos,
E. Calzetta, A. Feinstein,  
J. Iba{\~n}ez and A. Van Tonder for helpful
discussions. I am also grateful to the Physics Department of Brown
University for their hospitality and to the Grup de F\'{\i}sica
Te\`orica (IFAE) de l'Universitat Aut\`onoma de Barcelona.
This work has been partially supported by 
NSF grant PHY 95-14726 to The University of Chicago and by the Direcci\'o
General de Recerca de la
Generalitat de Catalunya through the grant 1995BEAI300165.

\appendix

\section{Useful  adiabatic expansions}

The coefficients for the Hadamard function in (\ref{eq:G(1)fR}), using
for simplicity the notation $z=\sin{t}$, are:
 
\bban {\bar
A}&=&{1\over 60\, L_1L_2\, \pi ^2},
\\
& &
\\
{\bar B}&=&
{525-426\, z^2+485\, z^4\over
60480\, (L_1L_2)^2 \, \pi ^2\, (1-z)^2(1+z)^2}
\\
& &
\\
C_{{\bar z}{\bar z}}&=&
{3597+2002\, z^2+3107\, z^4\over
332640\, L_1L_2\, \pi ^2\, (1-z)^2(1+z)^2}
\\
& &
\\
C_{{\bar x}{\bar x}}&=&
{561+2222\, z^2+1524\, z^4\over
166320\,  (L_1L_2)^2\, \pi ^2\, (1-z)(1+z)}
\\
& &
\\
C_{{\bar y}{\bar y}}&=&{1\over L_1L_2}\, C_{{\bar z}{\bar z}}
\\
& &
\\
D_{{\bar z}{\bar z}{\bar z}{\bar z}}&=&
{4433+12974\, z^2+3513\, z^4
\over 823680\, \pi ^2\, (1-z)^2(1+z)^2}
\\
& &
\\
D_{{\bar x}{\bar x}{\bar x}{\bar x}}&=&
{5005+31200\, z^2+45867\, z^4
\over 4324320\, (L_1L_2)^2\, \pi ^2}
\\
& &
\\
D_{{\bar y}{\bar y}{\bar y}{\bar y}}&=&
{1\over (L_1L_2)^2}\, D_{{\bar z}{\bar z}{\bar z}{\bar z}}
\\
& &
\\
D_{{\bar z}{\bar z}{\bar x}{\bar x}}&=&
{1001+4030\, z^2+2848\, z^4
\over 864864\, L_1L_2\, \pi ^2\, (1-z)(1+z)}
\\
& &
\\
D_{{\bar z}{\bar z}{\bar y}{\bar y}}&=&
-{143+260\, z^2+120\, z^4
\over 20592\, L_1L_2\, \pi ^2\, (1-z)^2(1+z)^2}
\\
& &
\\
D_{{\bar x}{\bar x}{\bar y}{\bar y}}&=&
{1\over L_1L_2}\, D_{{\bar z}{\bar z}{\bar x}{\bar x}}
\eean

The coefficients for the {\em midpoint
expansion} of the locally constructed Hadamard function (\ref{eq:Hada1})
are:

\bban
a_1^{(0)}&=&-R\,\left(\xi-{1\over 6}\right),\;\;\;\;
{\Delta ^{(2)}}_{\mu\nu}={1\over 12}R_{\mu\nu},
\\
& &
\\
a_2^{(0)}&=&
{1\over 2}\left({1\over 6}-\xi\right)^2\, R^2
+{1\over 6}\left({1\over 5}-\xi\right)\, {R_{;\alpha}}^\alpha
-{1\over 180}\, R^{\alpha\beta} R_{\alpha\beta}
+{1\over180}\,R^{\alpha\beta\gamma\delta}R_{\alpha\beta\gamma\delta},
\\
& &
\\
{a_1^{(2)}}_{\mu\nu}&=&
{1\over 24}\left({1\over 10}-\xi\right)\, R_{;\mu\nu}
+{1\over 120}\, {R_{\mu\nu ;\alpha}}^\alpha
-{1\over 90}\, {R^\alpha}_\mu R_{\alpha\nu}+
\\
& &
\\
& & {1\over 180}\, R^{\alpha\beta}R_{\alpha\mu\beta\nu}
+{1\over 180}\, {R^{\alpha\beta\gamma}}_{\mu}\,  
R_{\alpha\beta\gamma\nu},
\\
& &
\\
{\Delta ^{(4)}}_{\mu\nu\rho\tau}&=&
 {3\over 160}\, R_{\mu\nu ;\rho\tau}
+{1\over 288}\, R_{\mu\nu}R_{\rho\tau}
+{1\over 360}\, {{{R^\alpha}_\mu}^\beta}_\nu  
R_{\alpha\rho\beta\tau}.
\eean

\vskip 1.25 truecm
 
{\Large{\bf Figure captions}}
 
\vskip 0.5 truecm
 
\noindent
{\bf Fig. 1}
The colliding wave space-time consists of two
approaching waves, regions II and III, in a flat background, region
IV, and an interaction region, region I. The two waves move in the
direction of two null coordinates $u$ and $v$.
The four space-time regions
are separated by the two null wave fronts $u=0$ and $v=0$. The
boundary between regions I and II is  $\{0\leq u<\pi /2,\; v=0\}$, the
boundary between regions I and III is $\{u=0,\; 0\leq v <\pi /2\}$, and
the boundary of regions II and III with region IV is
$\Sigma =\{u\leq 0,\;v=0\}\cup\{u=0,\;v\leq 0\}$. Region I meets region IV
only at the surface $u=v=0$. The Killing-Cauchy horizon in the region
I corresponds to the hypersurface $u+v=\pi /2$ and plane wave regions
II and III meet such a Killing-Cauchy horizon
only at ${\cal P}=\{u=\pi /2,\; v=0\}$ and
${\cal P}'=\{u=0,\; v=\pi /2\}$ respectively. 
The hypersurfaces  $u=\pi /2$ in region II  and $v=\pi /2$ 
in region III are a type of topological singularities commonly referred as
folding singularities and they
must be identified with $\cal P$ and ${\cal P}'$ respectively.

\vskip 0.5 truecm

\noindent
{\bf Fig. 2}
The subset of Cauchy data which affects the evolution of the 
quantum field along the center $u=v$ of the plane wave  collision lies on
the segments 
${\bar\Sigma}_{\rm I}=\{0\leq u<\pi /4,\; v=0\}\cup\{u=0,\; 0\leq v <\pi /4\}$.
Region $\cal S$ is the causal future of this Cauchy data (or
equivalently, the causal past of the collision center).

We change the {\em mode propagation problem} for the plane wave
collision, in the causal past of the collision center, region $\cal S$,
by a much simpler Schr\"odinger-type problem which consists
in eliminating the dependence of the field equation on coordinate
$z$ by taking $z=0$, and substituting the Cauchy data which
come from the single plane wave regions on segments
${\bar\Sigma}_{\rm I}$
by much simpler Minkowski Cauchy data. This procedure is essentially
equivalent to modifying the space-time geometry in the causal
past of the collision center by eliminating
the dependence on coordinate $z$ in the
line element, setting $z=0$, and smoothly matching this
line element, through plane wave regions II and III, 
with the flat spacetime below the segment
$\{{t}=0,\; -\pi /4< z < \pi /4\}$.

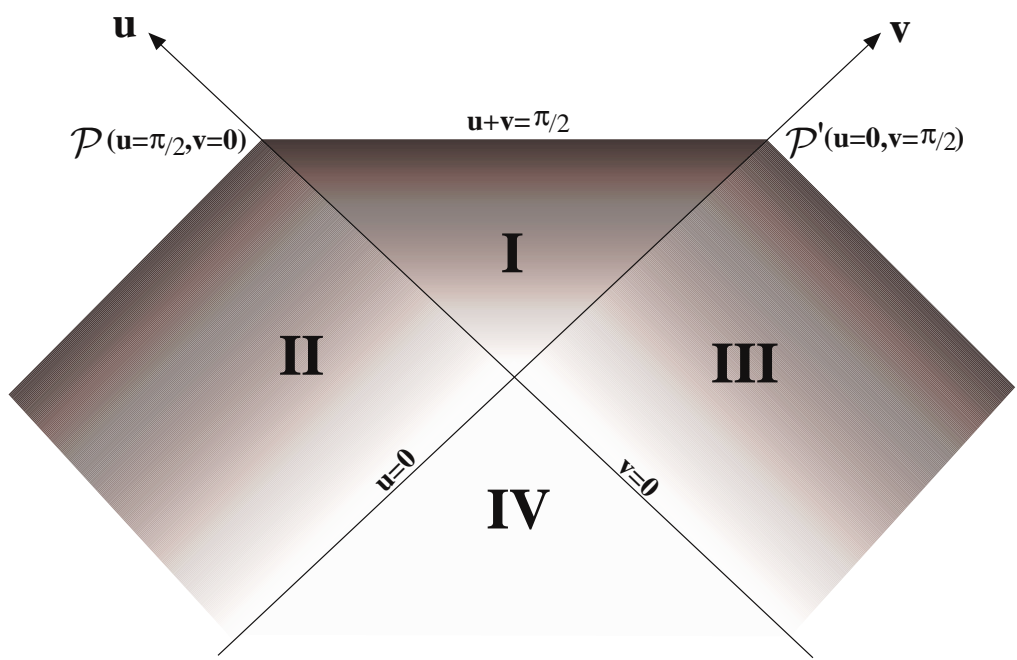
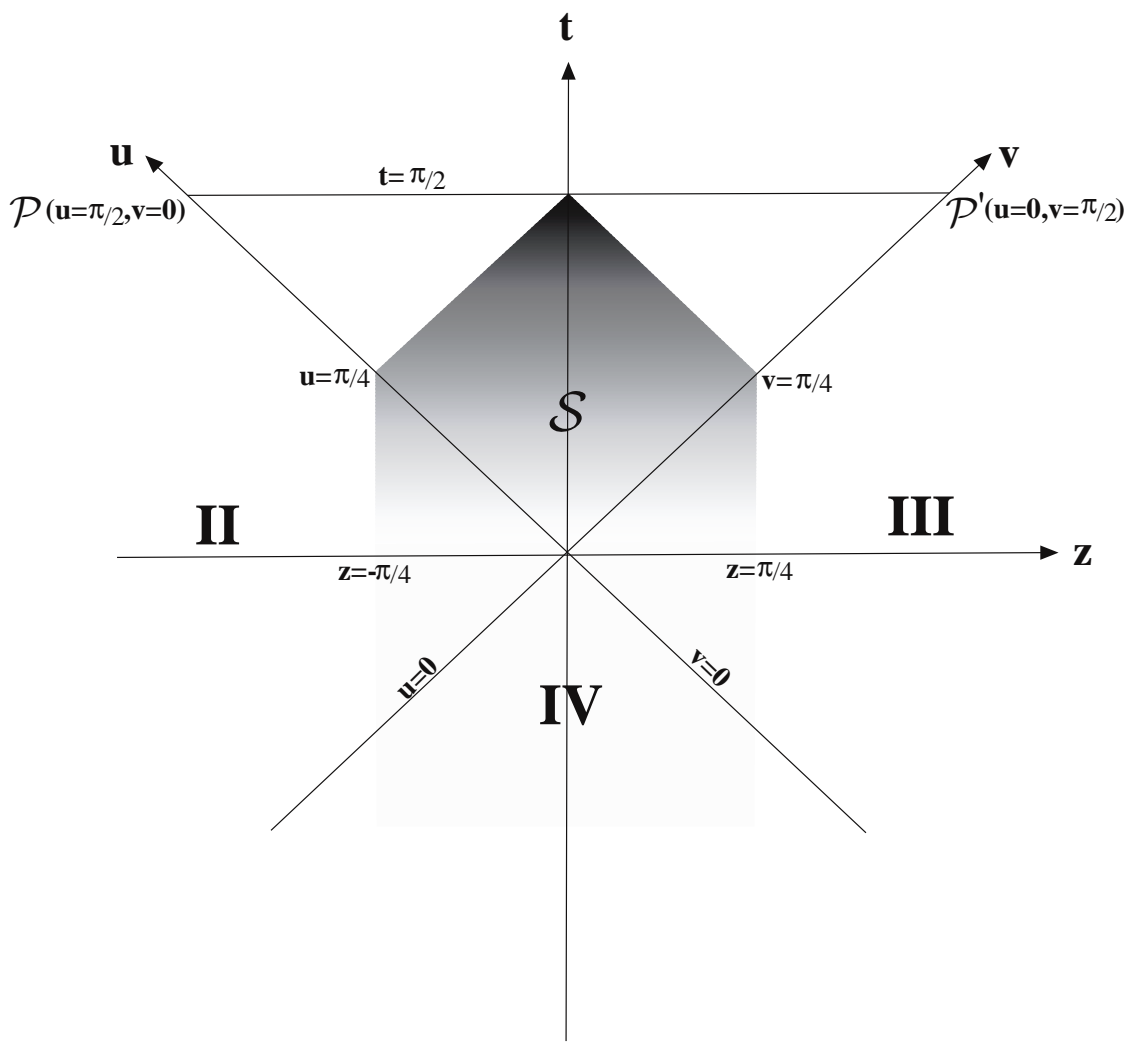
\end{document}